\documentclass[9pt,sigconf,nonacm,screen]{acmart}

\usepackage[all]{nowidow}
\usepackage[dvipsnames]{xcolor}
\usepackage{subcaption}
\usepackage{hyperref}
\usepackage{acronym}
\usepackage{xspace}

\usepackage[capitalise,noabbrev]{cleveref}
\crefformat{section}{\S#2#1#3}
\crefformat{subsection}{\S#2#1#3}
\crefformat{subsubsection}{\S#2#1#3}
\crefrangeformat{section}{\S#3#1#4 to~\S#5#2#6}
\crefrangeformat{subsection}{\S#3#1#4 to~\S#5#2#6}
\crefrangeformat{subsubsection}{\S#3#1#4 to~\S#5#2#6}
\newcommand{\name}{\emph{Provuse}}

\begin{document}

\author{Niklas Kowallik}
\affiliation{%
    \institution{TU Berlin}
    \city{Berlin}
    \country{Germany}}
\email{nk@3s.tu-berlin.de}
\orcid{0009-0006-9839-1864}

\author{Natalie Carl}
\affiliation{%
    \institution{TU Berlin}
    \city{Berlin}
    \country{Germany}}
\email{nc@3s.tu-berlin.de}
\orcid{0009-0000-5991-9255}

\author{Leon Pöllinger}
\affiliation{%
    \institution{TU Berlin}
    \city{Berlin}
    \country{Germany}}
\email{lp@3s.tu-berlin.de}
\orcid{0009-0001-1074-8624}

\author{Wei Wang}
\affiliation{%
    \institution{Huawei}
    \city{Berlin}
    \country{Germany}}
\email{weiwang2@huawei.com}
\orcid{0009-0005-9512-7884}

\author{Sharan Santhanam}
\affiliation{%
    \institution{Huawei}
    \city{Berlin}
    \country{Germany}}
\email{sharan.santhanam@huawei.com}
\orcid{0000-0002-9524-3253}

\author{David Bermbach}
\affiliation{%
    \institution{TU Berlin}
    \city{Berlin}
    \country{Germany}}
\email{db@3s.tu-berlin.de}
\orcid{0000-0002-7524-3256}

\title{\textsc{Provuse}: Platform-Side Function Fusion for Performance and Efficiency in FaaS Environments}

\keywords{cloud orchestration, FaaS, function, Function as a service, fusion, optimization, serverless computing}

\copyrightyear{2026}

\begin{abstract}
    Function-as-a-Service (FaaS) platforms provide scalable and cost-efficient execution but suffer from increased latency and resource overheads in complex applications comprising multiple functions, particularly due to double billing when functions call each other.
    This paper presents \name, a transparent, platform-side optimization that automatically performs function fusion at runtime for independently deployed functions, thereby eliminating redundant function instances.
    This approach reduces both cost and latency without requiring users to change any code.
    \name targets provider-managed FaaS platforms that retain control over function entry points and deployment artifacts, enabling transparent, runtime execution consolidation without developer intervention.

    We provide two implementations for this approach using the tinyFaaS platform as well as Kubernetes, demonstrating compatibility with container orchestration frameworks.
    An evaluation shows consistent improvements, achieving an average end-to-end latency reduction of 26.33\% and a mean RAM usage reduction of 53.57\%.

    These results indicate that automatic function fusion is an effective platform-side strategy for reducing latency and RAM consumption in composed FaaS applications, highlighting the potential of transparent infrastructure-level optimizations in serverless systems.
\end{abstract}

\maketitle

\section{Introduction}
\label{introduction}
The paradigm shift toward Serverless Computing, more specifically Function-as-a-Service (FaaS), has redefined cloud-native application development by abstracting infrastructure management away from the end-user~\cite{Hendrickson_Sturdevant_Harter_Venkataramani_Arpaci-Dusseau_Arpaci-Dusseau_2016,bermbach2021future}.
In FaaS applications, developers deploy discrete, ephemeral code snippets (functions) that are executed in response to incoming events.
These events either require an immediate, synchronous response or are triggered asynchronously without requiring a response.
While this offers unparalleled benefits in terms of auto-scaling and pay-as-you-go pricing, it introduces significant systemic overheads with regard to communication, invocation and runtime complexity~\cite{Jia_Witchel_2021,Qi_Monis_Zeng_Wang_Ramakrishnan_2022}.
Specifically, these overheads stem from the complexity of the underlying control plane required to provide a simplified user interface (e.g., scheduling, load balancing, and resource isolation), the additional latency and resource consumption introduced by function invocation and inter-function communication.

\begin{figure}
    \centering
    \includegraphics[trim={.5cm .4cm .5cm .2cm},clip,width=\linewidth]{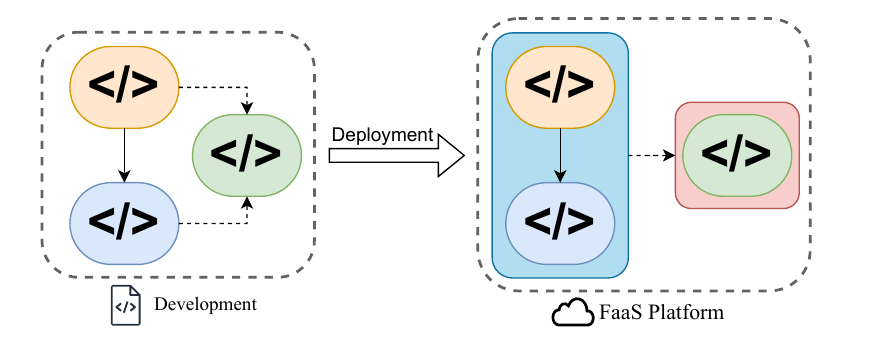}
    \caption{
        Function Fusion optimizes FaaS applications for latency and resource efficiency by colocating developer-defined functions across fewer runtimes, allowing function calls to be inlined rather than executed via remote invocation.
        }
    \label{fig1}
\end{figure}

A constant challenge in FaaS platforms is optimizing for both resource efficiency of the platform itself, reducing cost for users, and execution latency, which is required for low-latency real-time applications~\cite{eismann2021state}.
While application-level optimizations, such as code minification or selecting a faster programming language~\cite{mahmoudi2020placement,chadha2021architecture_specific}, can improve performance, platform-side optimizations allow improving existing applications without modifying user code and support additional optimization techniques~\cite{li2025hotswap}.
These are internal infrastructure enhancements, ranging from sandbox isolation techniques~\cite{moebius2024unikernel} to sophisticated scheduling algorithms, implemented by the provider~\cite{schirmer2023profaastinate, Tang2025XFaaS}.

Another approach to increase application performance is function fusion~\cite{schirmer2024fusionizepp}, which decouples the code that the users write from the function that is deployed on the platform itself.
By analyzing the call patterns between user code during the execution of the application, the code snippets that call each other frequently can be "fused" together into a single execution unit.
This eliminates the overhead of remote function calls as well as the \emph{double billing} problem~\cite{baldini2017trilemma}, while keeping the platform easy to use for users.

While previous work has shown the feasibility of function fusion by implementing it application-side~\cite{schirmer2024fusionizepp}, this paper focuses on the additional benefits that can be achieved by providing it as a platform-side offer.
By shifting optimization responsibilities to the platform, developers no longer need to manage deployment concerns.
Consequently, platform-side optimizations represent a logical progression from client-controlled approaches, enabling improvements that are fully transparent to the developer.

\section{Background and Related Work}
\label{background_related_work}

Function-as-a-Service (FaaS) is a cloud execution model in which providers manage infrastructure provisioning, scaling, and execution, while users supply application logic as discrete functions~\cite{shafiei_serverless_2022}.
This abstraction enables elasticity and fine-grained, pay-per-invocation billing, but also introduces performance variability and cost inefficiencies that motivate extensive platform-level optimization research~\cite{jonas_cloud_2019,serverless_survey_2021}.

\subsection{FaaS Execution Model and Platforms}

Upon invocation, a FaaS platform schedules a function, provisions or reuses an isolated execution environment, loads code and dependencies, initializes the language runtime, and executes the function logic.
Isolation is commonly implemented using containers or lightweight virtual machines, enabling multi-tenant execution while preserving fault and security boundaries.

FaaS platforms primarily differ in their approaches to scheduling, isolation, scaling, and resource management.
Lightweight runtimes, such as \emph{tinyFaaS}~\cite{pfandzelter2020tinyfaas} are designed for minimal overhead and fast function dispatch, making them suitable for experimental setups and edge deployments.
In contrast, \emph{Kubernetes}-based FaaS platforms (e.g., Knative~\footnote{\url{https://knative.dev/}}, OpenFaaS~\footnote{\url{https://www.openfaas.com/}}) build on container orchestration to provide declarative deployment, autoscaling, and integration with cloud-native ecosystems, at the cost of additional architectural complexity and runtime overhead~\cite{Kaviani_Kalinin_Maximilien_2019}.

These designs illustrate a spectrum ranging from minimal, specialized runtimes to feature-rich, production-grade orchestration frameworks, each making different trade-offs between performance, flexibility, and operational complexity.

\begin{figure}
    \centering
    \includegraphics[trim={.7cm .5cm .7cm .3cm},clip,width=\linewidth]{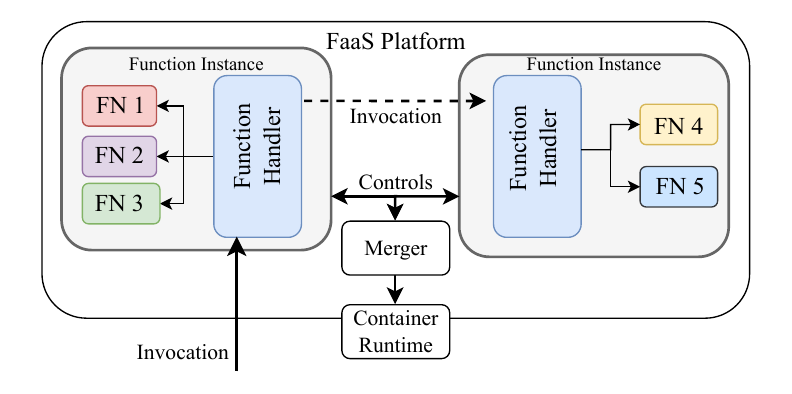}
    \caption{
        Design of the optimized FaaS platform.
        The \emph{Function Handler} is deployed together with function code, waiting for requests to invoke functions and monitoring outgoing requests to invoke the merging process.
        The Merger interacts with the container runtime to extract file systems and build new function instances hosting the code of multiple functions.
    }
    \label{fig:platform}
\end{figure}

\subsection{Platform-Side Optimization Strategies}

FaaS platforms rely on multi-tenancy to achieve economic efficiency, requiring isolation mechanisms that balance startup latency, resource overhead, and security guarantees~\cite{varghese2021survey,moebius2024unikernel}.
Isolation design directly affects scheduling flexibility, memory utilization, and execution environment reuse, thereby shaping the optimization space available to the provider.

Platform-side optimizations commonly target three complementary aspects.
First, sandbox management techniques, such as execution environment reuse, pre-initialization, and warm instance pooling, reduce cold-start frequency as well as runtime initialization costs.
Second, scheduling strategies exploit invocation affinity, historical execution behavior, and locality to minimize scheduling overhead and improve utilization.
Schedulers such as Hermes and Archipelago~\cite{Yan_Gao_Wu_Zhang_Hua_Huang_2021,Singhvi_Houck_Balasubramanian_Shaikh_Venkataraman_Akella_2019}, as well as more recent pull-based designs~\cite{hiku_2024}, represent different points in the design space, each making distinct trade-offs between placement flexibility, latency, and resource efficiency~\cite{placement_survey_2024}.
Third, despite the stateless programming abstraction, platforms employ low-latency mechanisms for ephemeral state handling to reduce external storage access in tightly coupled invocation patterns~\cite{klimovic_understanding_2018,sreekanti2020cloudburst}.

These optimizations are transparently controlled by the provider and require no changes to user code, making them well suited for systematic platform-level evaluation.

\subsection{Deployment Model and Cost Semantics}

This work focuses on the \emph{bring-your-own function code} deployment model, in which users submit source- or byte code.
The platform assumes responsibility for dependency resolution, environment preparation, and lifecycle management, maximizing its ability to apply provider-controlled optimizations.

A defining economic characteristic of FaaS platforms is fine-grained billing based on execution time and allocated resources~\cite{manner2023structured}.
In composed applications, this model can lead to \emph{double billing}, where synchronous function invocations incur redundant charges due to orchestration overhead and idle wait times~\cite{baldini2017trilemma,eismann_predicting_2020}.
This effect increases both latency and cost in serverless workflows, motivating techniques that reduce redundant execution boundaries~\cite{eismann_review_2021}.

\subsection{Function Fusion and Execution Consolidation}

Function fusion is one such execution consolidation technique.
It combines multiple logically distinct functions, originally written and deployed as independent units, into a single execution unit~\cite{Elgamal_Sandur_Nahrstedt_Agha_2018,schirmer2022fusionize}.
By eliminating redundant scheduling, runtime initialization, and inter-function communication, fusion preserves program semantics while reducing overhead.

Earlier work on function fusion emphasized user-controlled decisions, such as explicit annotations, configuration parameters, or deployment-time hints, often coupled with specialized programming or runtime models.
More recent works, including by Schirmer et al.~\cite{schirmer2024fusionizepp}, and Kowallik et al.~\cite{kowallik2026konflux}, automatically derive fusion decisions from empirical invocation traces to optimize latency and cost.
Other approaches explore offline analysis, static deployment, or non-standard runtimes~\cite{liu_faaslight_2023,agache_firecracker_nodate}.

Empirical studies show that a substantial fraction of FaaS workloads exhibit synchronous invocation patterns, where one function immediately triggers another and blocks on its result~\cite{schirmer2024fusionizepp,kowallik2026konflux}.
In such cases, strict execution separation introduces avoidable overhead, making function fusion a particularly effective optimization strategy.

\begin{figure*}
    \centering
    \begin{minipage}{0.47\textwidth}
            \centering
            \includegraphics[trim={.5cm .5cm .5cm .5cm},clip,width=\linewidth]{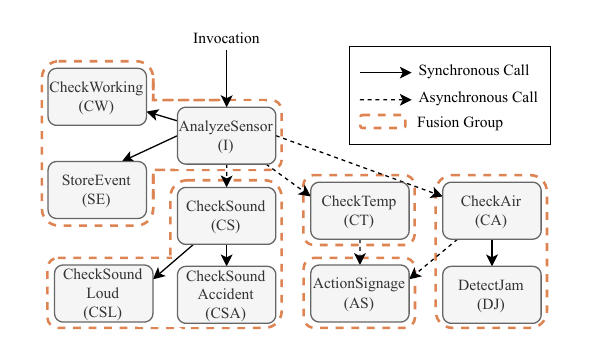}
            \caption{
        Call graph of the Fusionize++ IoT application \cite{schirmer2024fusionizepp}. Solid edges denote synchronous invocations, dashed edges asynchronous execution.
        The workflow starts at AnalyzeSensor (I), combining sequential steps with parallel analysis of temperature, air quality, and traffic.
        Dashed shapes indicate the theoretical fusion groups that would result from fusing synchronous calls.
        }
        \label{fig:iot}
    \end{minipage}%
    \hfill
    \begin{minipage}{0.47\textwidth}
            \centering
    \includegraphics[width=0.55\linewidth,trim={.5cm .7cm .5cm .7cm},clip]{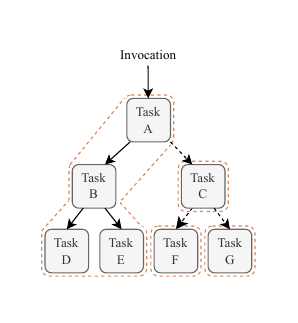}
    \caption{
        Call graph of the TREE application from Fusionize++~\cite{schirmer2024fusionizepp}, highlighting theoretical fusion groups based on synchronous calls with a dashed line.
        A synchronously invokes B, which calls D and E, while A also triggers an asynchronous branch via C to F and G.
        The asynchronous path dominates the workload, requiring far more computation than the synchronous branch.
        }
    \label{fig:tree}
    \end{minipage}
\end{figure*}

\section{Design}
\label{design}
We propose \name, a design that extends existing FaaS environments.
The design introduces two tightly coupled components, the \emph{Merger} and the \emph{Function Handler}.
Together, these components enable platform-side optimizations by transparently applying function fusion at runtime while preserving the logical function boundaries exposed to developers.

\cref{fig:platform} illustrates the high-level architecture and the interaction between these components and deployed function instances.

The \emph{Function Handler} (see \cref{fig:platform}) actively coordinates requests and invokes functions.
It orchestrates all function invocations by dispatching inbound requests to the appropriate local function.
To support function fusion, the Function Handler also detects synchronous communication and notifies the Merger when it should combine function instances.
Although network traffic monitoring can reveal such communication patterns, the \emph{bring-your-own-function-code} model allows the platform to control each function’s entry point and directly access the sockets of every function instance.
The Function Handler therefore observes system-level outbound socket connections initiated by a function and determines whether each socket operates in blocking (synchronous) mode, in which the issuing thread waits for completion, or in non-blocking (asynchronous) mode, in which execution continues while the request remains in flight.
When the Function Handler detects synchronous communication, it triggers the Merger by submitting a fusion request that includes the relevant function identifiers, such as function names, IP addresses, or port numbers.

The Merger actively consolidates multiple independently deployed functions into a single container.
After receiving a fusion request that specifies two active function instances, the Merger creates a new function image based on the referenced containers.
It exports their container file systems, merges them into a unified file system, and builds a new container image from the merged state.
The platform enables this process by controlling both the function container build process and the storage layout of function code used by the Function Handler.
As a result, the Merger can recreate the original directory structure and copy the code of both function instances behind a new Function Handler with minimal effort.

To prevent file overwrites caused by colliding function names, the Merger preserves the original identifiers of each function instance while copying them into the shared file system.
After assembling the combined file system, the Merger builds a new container image and deploys the resulting container.
The platform monitors the new container until all health checks succeed and then routes incoming requests for the local functions to the combined instance.
Once traffic has been fully redirected, the platform terminates the original containers to free resources.
This process improves resource utilization while preserving function addressability and performance.

\section{Implementation}
\label{implementation}

We implemented \name{} on both tinyFaaS~\cite{pfandzelter2020tinyfaas}, a small-scale state-of-the-art edge-based Function-as-a-Service platform, and Kubernetes\footnote{\url{https://kubernetes.io/}}\!\!, a container orchestration platform that is used as a basis for many FaaS-platforms such as Knative~\cite{Wen_Chen_Jin_Liu_2023}.

The \emph{Function Handler} component is implemented in Python, a widely used language for serverless applications~\cite{eismann_review_2021}.
When a Function Handler receives a function call, a dedicated thread monitors the state of its connection sockets.
If a socket communicates with an IP address within the FaaS platform and is in a blocking state, causing the associated thread to wait, the Function Handler dispatches a request to the Merger component.
This request includes the name of the function instance hosting the Handler and the IP address and port of the called function instance, which uniquely identifies the remote function on both platforms.

The \emph{Merger} is implemented in Rust and packaged as a container image for deployment either alongside FaaS platform components or, in Kubernetes environments, alongside function instance pods.
Upon invocation, the Merger resolves the remote identifier provided by the Function Handler to locate the relevant function instances.
It then extracts, combines, and deploys a merged version of the two functions using container runtime access.
Once the new combined function instance is running and healthy, incoming request traffic is redirected to it.
The function instances that where combined are stopped and deleted, as soon as they are no longer processing requests.

Although the current prototype of \name{} supports only Python as a language for function code, all platform-side mechanisms are language-agnostic, allowing straightforward extension to additional programming languages for function implementation.

For Kubernetes-based platforms, the function fusion mechanism leverages existing primitives for deployment and lifecycle management.
Access to individual function instances can be maintained and updated via Kubernetes services, facilitating smooth integration with the platform.
For tinyFaaS the combined function instance overwrites the old function entries in the API gateway of tinyFaaS, to reroute new requests to the newly deployed and fused function instance.
\section{Evaluation}
\label{evaluation}

\begin{figure}
    \centering
    \includegraphics[width=\linewidth]{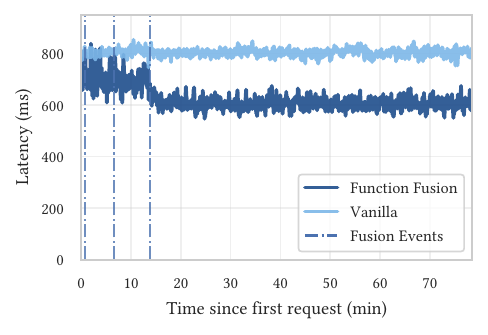}
    \caption{
        Time series of end-to-end latency for the IOT application on tinyFaaS.
        The vanilla deployment consistently exhibits higher latencies, while the function fusion deployment achieves a 28.9\% latency reduction.
        The vertical lines represent the finished merge event, where a new combined function instance is healthy and running.
    }
    \label{img:iot:tinyfaas}
\end{figure}

In earlier sections, we argue that \name{} improves performance and resource efficiency without increasing developer-facing complexity.
Because all fusion decisions are performed at the platform level and require no changes to application code or deployment configuration, developer interaction remains identical to that of a vanilla FaaS deployment.
Consequently, developer usability is treated as a design property rather than an empirical metric, and our evaluation focuses exclusively on measurable performance and resource efficiency effects.

\subsection{Experiments Setup}
\label{experiments}
To this end, we evaluate the performance impact of the proposed function merging approach through systematic benchmarking experiments.
We base our evaluation on two FaaS applications previously used by Schirmer et al.~\cite{schirmer2022fusionize, schirmer2024fusionizepp} to demonstrate function fusion effects: \emph{TREE} and \emph{IOT}.
The TREE application (see \cref{fig:tree}) represents a minimal use case for function fusion, with a call graph comprising a simple binary tree with synchronous calls on one side and asynchronous calls on the other side (see \cref{fig:tree}).
The IOT application (see \cref{fig:iot}) represents a more realistic use case where input from IoT sensors is analyzed and stored (see \cref{fig:iot}).

We run all experiments on two virtual machines: one hosting the system under test (SUT) and one running a benchmarking client~\cite{bermbach2017book}.
The VMs run on QEMU/KVM with 4 vCPUs and 16~GB RAM each.
During experiments, the benchmarking client VM is hosted on a physically separate server to the SUT VM with a 10~Gbit/s network connection between both servers.

Each experiment consists of 10,000 HTTP requests, generated by k6\footnote{\href{https://k6.io/}{https://k6.io/}} under constant request rates of 5 requests per second, which is the rate also used in previous function fusion papers~\cite{schirmer2024fusionizepp,kowallik2026konflux}.
The experiments are conducted in both tinyFaaS and Kubernetes environments, with two benchmarking runs per deployed application: One with the merging mechanism enabled and one with it disabled.

This setup supports a controlled and direct comparison of execution latency and system behavior between the automatic platform-side effects of function fusion and the vanilla FaaS applications.

\subsection{Results}
\label{results}

\begin{figure*}
\centering
\includegraphics[width=\textwidth]{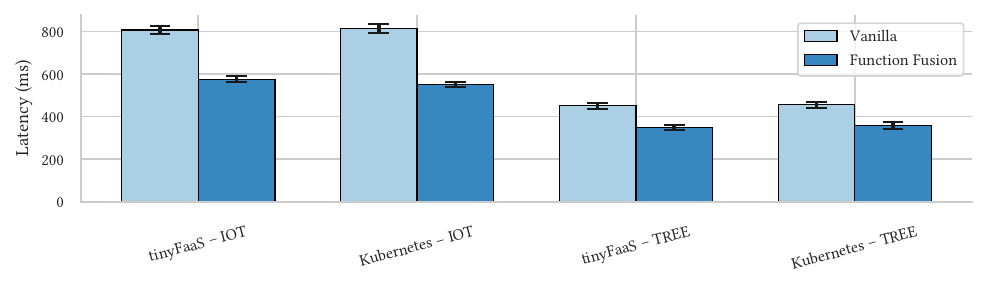}
\caption{
    Median end-to-end latency (in milliseconds) for vanilla and function fusion deployments across tinyFaaS and Kubernetes for the IOT and TREE applications.
    Whiskers indicate narrow latency ranges, reflecting stable performance.
    The function fusion configuration achieves on average 26.3\% lower median latency across all benchmarks.
}
\label{img:bench:bar}
\end{figure*}

We evaluate our approach using two representative applications (TREE and IOT) deployed on two distinct serverless platforms (tinyFaaS and Kubernetes).
For each configuration, we measure end-to-end request latency and platform-level resource usage in the form of RAM usage, comparing a baseline \emph{vanilla} deployment with our optimized \emph{function fusion} deployment.

The time series of the IOT application running on tinyFaaS, shown in \cref{img:iot:tinyfaas}, provides a representative overview of the behavior observed across all conducted experiments.
During the initial phase, both deployment variants exhibit comparable performance characteristics.
However, request latency decreases progressively with each completed merge process, and after the optimization phase concludes, a clear performance improvement is visible.
Specifically, the \emph{vanilla} deployment consistently incurs higher request latency, whereas the \emph{function fusion} deployment achieves notably lower response times.

Quantitatively, the IOT application experiences a reduction in median latency from 807\,ms to 574\,ms, corresponding to an improvement of 28.9\%, while simultaneously reducing RAM usage by approximately 57\%.
Similarly, for the TREE application, the median latency decreases from 452\,ms to 350\,ms (22.6\%), accompanied by a 50\% reduction in RAM consumption.

Both time series show short-term fluctuations typical of FaaS execution environments; however, their overall distributions remain stable.
The remaining time series for other application-platform combinations exhibit comparable trends and are omitted for brevity.

All Kubernetes experiments exhibit similar improvements.
For IOT, median latency decreases from 815\,ms to 551\,ms (32.4\%), while TREE latency is reduced from 456\,ms to 358\,ms (21.5\%).
RAM usage reductions closely match those observed on tinyFaaS, at approximately 57\% for IOT and approximately 50\% for TREE.

\cref{img:bench:bar} summarizes median end-to-end latency across all four experimental configurations.
Across all benchmarks, the Function Fusion deployment consistently outperforms the vanilla deployment, achieving an average median latency reduction of 26.3\%.
Together, these results demonstrate that automatic platform-side function fusion yields consistent latency and efficiency improvements across both platforms and application types.

Overall, provider-side function fusion reduces median latency by 26.3\% and RAM usage by 53.6\% on average across all evaluated experiments.
\section{Discussion \& Future Work}
\label{discussion}

Platform-side mitigation of double billing through function fusion in composed FaaS applications is both feasible and effective.
By fusing synchronously composed functions into a single execution unit, \name{} mitigates redundant billing effects that arise from chained invocations in fine-grained FaaS pricing models~\cite{baldini2017serverless}.

The observed reductions in end-to-end latency stem primarily from avoiding redundant function invocations and associated scheduling overheads.
At the same time, consolidating execution reduces the number of concurrently allocated function instances, directly translating into lower platform RAM consumption.
Importantly, these benefits are consistent across both lightweight (tinyFaaS) and cloud-level (Kubernetes) environments, as well as across different application characteristics.

Beyond mitigating the effects of double billing, the results highlight the broader potential of transparent, platform-driven optimization mechanisms in serverless systems.
Techniques such as peak shaving~\cite{schirmer2023profaastinate,Tang2025XFaaS} or instance pre-warming to absorb bursty workloads~\cite{Verma_Goel_Rani_2024} could be integrated using similar principles, further improving latency and resource efficiency without imposing additional burdens on developers.

Nevertheless, certain limitations apply.
Function fusion necessarily reduces isolation between fused functions by colocating them within a shared execution environment.
As a result, \name{} assumes that fusion decisions are restricted to functions belonging to the same trust domain.
The overhead incurred by function fusion, including container image reconstruction and redeployment, is amortized over subsequent invocations of the fused execution unit.

Applications that are already highly optimized, particularly fully asynchronous or non-blocking workloads, may see limited to no benefit from execution consolidation, as latency and resource savings depend strongly on application structure.
Additionally, our evaluation focuses on the \emph{bring-your-own-function-code} model, in which our platform does not analyze the functions' code and solely optimizes at the invocation level.
This limits developers in the applications they can deploy, as third-party containers (``bring-your-own-container'') currently are not part of our prototype.

\label{future_work}
However, future extensions could aim to relax this limitation by allowing user-defined container environments while preserving the platform's ability to perform global, automated optimizations.
Especially hybrid optimizations between the \emph{bring-your-own-function-code} and \emph{bring-your-own-container} model, are of interest for platform-side optimizations.
Further, our prototype implementation currently supports only Python-based function code.
This restriction is not inherent to the design but reflects the prototype's scope.

\section{Conclusion}
\label{conclusion}

Transparent, platform-side execution consolidation is a practical and effective strategy for improving the efficiency of composed FaaS applications.
By preventing double billing on the platform-side and reducing both latency and resource consumption, the proposed approach provides a concrete step towards more cost- and performance-efficient serverless platforms.
Future work may explore the integration of complementary optimization techniques and hybrid deployment models to extend these benefits to a broader range of serverless workloads.
We believe that further transparent improvements to FaaS platforms will be instrumental in reducing developer burden and enabling more efficient application development.
\balance

\bibliographystyle{ACM-Reference-Format}
\bibliography{bibliography.bib}

\end{document}